
\input phyzzx
\voffset = -0.4in
\footline={\ifnum\pageno=1 \nulline \else\newfootline \fi}
\def\nulline{{\hfill}}
\def\newfootline{\advance\pageno by -1\hss\tenrm\folio\hss}
\rightline {June 1994}
\rightline {QMW--TH--94/28}
\title {TARGET SPACE DUALITY  IN ORBIFOLDS \break
 WITH CONTINUOUS AND DISCRETE WILSON LINES.\break}
\author{D. Bailin$^{a *}$, \ A. Love$^{b}$,  \ W. A. Sabra
$^{b **}$\ and \ S. Thomas$^{c ***}$}
\address {$^{a}$School of Mathematical and Physical
Sciences,\break
University of Sussex, \break Brighton U.K.}
\address {$^{b}$Department of Physics,\break
Royal Holloway and Bedford New College,\break
University of London,\break
Egham, Surrey, U.K.}
\address {$^{c}$
Department of Physics,\break
Queen Mary and Westfield College,\break
University of London,\break
Mile End Road, London,  U.K.}
\footnote*{e-mail address: D.Bailin@SUSSEX.AC.UK.}
\footnote{**}{e-mail address: UHAP012@VAX.RHBNC.AC.UK.}
\footnote{***}{e-mail address: S.THOMAS@QMW.AC.UK.}
\abstract {Duality symmetry is studied for heterotic string orbifold
compactifications in the presence of a general background which in addition to
the metric and antisymmetric tensor fields contains both discrete and
continuous Wilson lines. }
\REF\one{ L. Dixon, J. Harvey, C. Vafa, and E. Witten,
 {\it Nucl. Phys.} {\bf B261} (1985) 678,  {\bf B274} (1986) 285;
K. Narain, M. Sarmadi and C. Vafa, {\it Nucl. Phys.} {\bf B288} (1987) 551,
 {\it  B356} (1991) 163.}
\REF\two{L. E. Ib\'{a}\~{n}ez, H. P. Nilles and F. Quevedo,
 {\it Phys. Lett.} {\bf B187} (1987) 25.}
\REF\three{L. E. Ib\'{a}\~{n}ez, H. P. Nilles and F. Quevedo,
 {\it Phys. Lett.} {\bf B192} (1987) 332.}

\REF\four{ L. E. Ib\'{a}\~{n}ez, J. Mas, H. P.  Nilles
and F. Quevedo, {\it Nucl.Phys.} {\bf B301}(1988) 157.}
\REF\five{ T. Mohaupt, {\it MS--TPI} 93--09.}
\REF\six{L. E. Ib\'{a}\~{n}ez, J. E. Kim, H. P. Nilles and F. Quevedo,
 {\it Phys. Lett.} {\bf B191} (1987) 3.}
\REF\seven{ S. Ferrara, C. Kounnas and M. Porrati,
{\it Phys. Lett.} {\bf B181} (1986) 263.}
\REF\eight{M. Cvetic, J. Louis and B. A. Ovrut, {\it Phys. Lett. B} {\bf 206}
(1988) 227.}
\REF\nine{G. L. Cardoso, D. Lust and T. Mohaupt, HUB-IEP-94/6.}
\REF\eleven{ D. Bailin, A. Love, W. A. Sabra and S. Thomas,
{\it Mod. Phys. Lett.} {\bf A}9 (1994) 1229, A. Love, W. A. Sabra and S.
Thomas, QMW--TH--94/05, to appear in {\it Nucl.Phys.} {\bf B}.}
\REF\twelve{R. Dijkgraaf, E. Verlinde and H. Verlinde, {\it Comm. Math.
Phys.}{\bf 115} (1988) 649.}
\REF\thirteen{ R. Dijkgraaf, E. Verlinde and H. Verlinde,
On Moduli Spaces of Conformal Field Theories with $c \geq 1$, Proceedings
Copenhagen Conference, Perspectives in String Theory,
edited by P. Di Vecchia and J. L. Petersen, World Scientific, Singapore, 1988.}
\REF\fourteen{ A. Shapere and F. Wilczek, {\it Nucl.Phys.} {\bf B320} (1989)
669.}
\REF\fifteen{ M. Dine, P. Huet and N. Seiberg,
{\it Nucl. Phys.} {\bf B322} (1989) 301.}
\REF\sixteen{ J. Lauer, J. Mas and H. P. Nilles, {\it Nucl. Phys.} {\bf B351}
(1991) 353.}
\REF\seventeen{ K. Kikkawa and M. Yamasaki, {\it Phys. Lett.} {\bf B149},
(1984) 357;
N. Sakai and I. Senda, {\it Prog. Theor. Phys.} {\bf 75} (1984) 692}
\REF\eighteen{A. Giveon, E. Rabinovici and G. Veneziano,
{\it Nucl. Phys.} {\bf B322} (1989) 167.}
\REF\nineteen{W. Lerche, D. L\"ust and N. P. Warner,
{\it Phys. Lett.} {\bf B231} (1989) 417.}
\REF\twenty{J. Erler and M. Spalinski, preprint MPI-PH-92-61,TUM-TH-147-92
(1992).}
\REF\twentyone{V. S. Kaplunovsky, {\it Nucl. Phys.} {\bf B307} (1988) 145}
\REF\twentytwo{L. J.
Dixon, V. S. Kaplunovsky and J. Louis,  {\it Nucl. Phys.} {\bf B355} (1991)
649.}
\REF\twentythree{J. P. Derendinger, S. Ferrara, C. Kounas and F. Zwirner, {\it
Nucl.
Phys.} {\bf B372} (1992) 145, {\it Phys. Lett.} {\bf B271} (1991) 307.}
\REF\twentyfour{A. Giveon, M. Porrati and E. Rabinovici,
RI-1-94, NYU-TH.94/01/01.}
\REF\twentyfive{ P. Mayr and S. Stieberger, {\it Nucl. Phys}. {\bf B407} (1993)
725.}
\REF\twentysix{ D. Bailin, A. Love, W. A. Sabra and S. Thomas,
{\it Phys. Lett.} {\bf B320} (1994) 21.}
\REF\twentyseven{ D. Bailin, A. Love, W. A. Sabra and S. Thomas,
{\it Mod. Phys. Lett.} {\bf A}9 (1994) 67.}
\REF\mns{ P. Mayr, H.P. Nilles and S. Stieberger,  {\it Phys. Lett.}
 {\bf B317} (1993) 53.}
\REF\twentyeight{K. S. Narain, {\it Phys. Lett.} {\bf B169} (1987) 41;
 K. S. Narain, M. H. Sarmadi  and E. Witten,  {\it Nucl. Phys.} {\bf B279
}(1987)
369.}
 \REF\twentynine{M. Spalinski, {\it Phys. Lett.} {\bf B275} (1992) 47;
J.  Erler, J.  Jungnickel, H. P. Nilles,  {\it Phys. Lett.} {\bf B276} (1992)
303.}
\REF\stef{ S. Stieberger, private communication and forthcoming preprint}

Orbifold compactified heterotic string theories [\one, \two, \three] provide
phenomenologically promising string compactified backgrounds as they give rise
to semi-realistic four dimensional quantum field theories [\two, \six]. The
orbifold models are characterized by a set of continuous parameters referred to
as moduli. These moduli are the marginal deformations of the underlying
conformal field theory of the orbifold. They  appear in  the space-time
supersymmetric four dimensional  Lagrangian with a flat potential to all orders
in perturbation theory.
The moduli space of the orbifold compactification is a subspace  of that of
the toroidal compactification obtained by demanding that the twist action on
the underlying Narain lattice is an automorphism.
The coset structure of the moduli space for the ${\bf Z}_3$ orbifold, in the
absence of Wilson lines, has been given in [\seven] and for general ${\bf Z}_N$
orbifolds in [\eight]. However,  in  [\eight], the symmetries of the conformal
field theory are employed to determine the Kahler potential which in turn fixes
the form of the coset structure of the moduli space.
In heterotic string theory, the moduli space of the orbifold is enlarged when
the gauge twist in the $E_8\times E_8$ root lattice is realized by a rotation.
The extra moduli are continuous Wilson lines [\three, \four, \five].
More recently, the coset structure of moduli space of the untwisted moduli
including continuous Wilson lines has been identified in [\nine].  Locally, the
untwisted moduli space is determined by the eigenvalues of the twist.

However, string theories have a novel discrete symmetry, the so-called  target
space duality [\twelve-\twenty] (see also [\twentyfour ] and references
therein), consisting of discrete reparametrizations of the background fields
(moduli) leaving the underlying conformal field theory invariant. Therefore,
the
physical moduli space has the form of an orbifold given as the quotient of the
moduli space by a discrete symmetry group.

In two-dimensional ${\bf Z}_2$-orbifold compactification, or in an eigenspace
of a six-dimensional orbifold where the twist has an eigenvalue  $-1$ and in
which the eigenspace lies entirely in a two-dimensional sublattice of the
orbifold six-dimensional lattice, the duality symmetry is given by $SL(2,
Z)_T\times SL(2, Z)_U,$ where $T$ and $U$ are the complex moduli parametrizing
the complex plane with the eigenvalue $-1$ [\thirteen].
This is the symmetry of the moduli-dependent threshold corrections to the gauge
coupling constants [\twentyone-\twentythree] in this case. However if the
complex plane does not lie entirely in a two-dimensional sublattice then the
duality group is broken down to a subgroup of  $SL(2, Z)_T\times SL(2, Z)_U$
[\twentyfive-\twentyseven]. Note that if the eigenvalue of the twist is
different from $-1$ the $U$ modulus is frozen to a constant phase factor and
the duality symmetry is given by the $T$-duality. Also it is known  that the
presence of discrete Wilson lines  break the duality group [\eleven,\twenty].
In [\eleven], the duality symmetry of the moduli-dependent threshold
corrections to the gauge coupling constants in the presence of discrete Wilson
lines background was determined.
Here we are interested in studying the duality symmetry of orbifold
compactified heterotic string theory in the presence of a general background
which contains both discrete and continuous Wilson lines. Of particular
interest are the symmetries of the twisted sectors  contributing to the
threshold corrections of the gauge coupling constants.
\footnote*{ Recent progress has been made concerning the calculation of
such threshold corrections in the presence of continuous Wilson lines [\stef ]
} Some consequences of discrete Wilson line
contributions to threshold corrections has been recently discussed in [\mns ] .

Consider a ten-dimensional $E_8\times E_8$ heterotic string compactified on a
$d$-dimensional torus  [\twentyeight] defined  as a quotient of
${\bf R}^d$ with respect to a lattice $ \Lambda$ defined by
$$\Lambda=\Big \{\sum_{i=1}^d a^ie_i,\qquad a^i\in Z\Big \}.\eqn\lah$$
The background is described by the metric $G_{ij}=e_i.e_j ,$ an antisymmetric
field $B_{ij}$ and Wilson lines $W_i^I,$ where the index $I$ refer to the gauge
degrees of freedom of the  $E_8\times E_8$ lattice.
The left and right momenta for the compactified string coordinates are given by
$$\eqalign{{\bf P}_L=&\Big({{\bf m}\over2} + ({\bf G - B}-{1\over4} {\bf W}^t
{\bf CW) {\bf n}} - {1\over2}{\bf W}^t{\bf  C l},\  {\bf l+W n}\Big)=\Big({\bf
p}_L,\ \tilde{{\bf p}}_L\Big),\cr {\bf P}_R=&\Big({{\bf p}\over2} - ({\bf  G+
B}+{1\over4} {\bf W}^t {\bf C W) n} - {1\over2}{\bf W }^t{\bf  C l},\  {\bf
0}\Big )=\Big({\bf p}_R,\ \bf 0\Big),}\eqn\mo$$
where $\bf n$ and $\bf m$ the windings and momenta respectively, are
$d$-dimensional integer valued vectors,  $\bf l$ is a 16-dimensional vector
representing the internal gauge quantum numbers and $\bf C$ is the Cartan
metric of the self-dual Euclidean root lattice of $E_8\times E_8$.

The left and right momenta of the compactified string coordinates
have the contribution $H$ and $S$ to the scaling dimension and spin
of the vertex operators of the underlying conformal field theory given by
$$\eqalign{H= &{1\over2} \Big({\bf p}^t_L {\bf G}^{-1} {\bf p}_L + \tilde{{\bf
p}}^t_L {\bf C} \tilde{{\bf p}}_L+{\bf p}^t_R {\bf G}^{-1} {\bf p}_R\Big),\cr
S= &{1\over2} \Big({\bf p}^t_L {\bf G}^{-1} {\bf p}_L + \tilde{{\bf p}}^t_L
{\bf C} \tilde{{\bf p}}_L-{\bf p}^t_R {\bf G}^{-1} {\bf p}_R\Big).}\eqn\ver$$
One way to study the duality transformations of the spectrum is to  write $H$
and $S$ in the following quadratic forms [\twentynine]
$$H={1\over2}  {\bf u}^t {\bf \Xi} {\bf u}, \qquad S={1\over2} {\bf u}^t \eta
{\bf u},\eqn\scale$$
where
$$\eqalign{{\bf u} =& \pmatrix{\bf n \cr \bf m\cr \bf l}, \quad  \eta =
\pmatrix{{\bf 0} & {\bf 1}_d &{\bf 0}\cr
	{\bf 1}_d & {\bf 0}&{\bf 0}\cr {\bf 0}&\bf 0&{\bf C} },\cr
\bf\Xi=&\pmatrix{
{1\over2}{\bf D}{\bf G}^{-1} {\bf D}^t &{\bf 1}_d +{1\over2}{\bf D}{\bf
G}^{-1}&-{1\over2}{\bf D}{\bf G}^{-1} {\bf W}^t {\bf C } \cr
 {\bf 1}_d+{1\over2}{\bf G}^{-1}{\bf D}^t &{1\over2}{\bf G}^{-1}
&-{1\over2}{\bf G}^{-1} {\bf W}^t{\bf C} \cr
-{1\over2}{\bf C}{\bf W}{\bf G}^{-1}{\bf D}^t&-{1\over2}{\bf CWG}^{-1} & {\bf
C}+{1\over2}{\bf C}{\bf W}{\bf G}^{-1}{\bf W}^t{\bf C}\cr} }\eqn\nicosia$$
with ${\bf D}=2\Big({\bf B}-{\bf G}-{1\over4}{\bf W}^t{\bf C}{\bf W}\Big)$ and
${\bf 1}_d$ is the $d\times d$ identity matrix.

The  discrete target space duality symmetries are then defined to be all
integer-valued linear transformations of the quantum numbers
${\bf n}, {\bf m} $ and ${\bf l} $
leaving the spectrum invariant.
Denote  these  linear transformations by ${\bf \Omega}$
and define their action on the quantum numbers as
$${\bf \Omega}:{\bf u} \longrightarrow {\bf\Omega}^{-1}{\bf u}.\eqn\lin$$
In order for these discrete transformation to preserve $S$, the transformation
matrix $\bf\Omega$ should satisfy the condition:
$${\bf\Omega}^{t}\eta\bf\Omega= \eta.\eqn\con$$
This means that ${\bf\Omega}$ is an element of $O(d+16,d;Z)$. Moreover,
requiring the  invariance of $H$ under the duality transformation
induces a   transformation on the moduli.
Such a transformation defines the action of the duality group  and is given by
$$
{\bf \Omega}:{\bf\Xi} \longrightarrow {\bf\Omega}^t \bf\Xi
\bf\Omega.\eqn\sand$$

The above analysis can be generalized to the orbifold case including both
discrete and continuous Wilson lines in the following manner. Consider the case
where the twist acts  on the orbifold and the $E_8\times E_8$ lattice vectors.
To define the action of the twist on the quantum numbers we demand that under
the action of the twist both ${\bf p}_{L}$ and ${\bf p}_{R}$ are rotated by
${\bf Q}^*={{\bf Q}^t}^{-1}$ and  $ \tilde{{\bf p}}_L$ is rotated by $\bf M,$
where ${\bf Q}$ is a $d\times d$ matrix defining the action of the twist on the
$d$-dimensional orbifold lattice, ${\bf M}$ is the action of the twist on the
$E_8\times E_8$ lattice satisfying ${\bf M}^t{\bf C}{\bf M}={\bf C}.$
Using the fact that the winding numbers transform as $\bf n\longrightarrow \bf
Qn$ then for $ \tilde{{\bf p}}_L$ to be rotated by $\bf M$ and such that  the
quantum numbers $\bf l$ are transformed as integers, the Wilson line must
satisfy
$$ {\bf MW-WQ}={\bf V}\in Z.\eqn\back$$
If we also demand that the background fields $\bf G$ and $\bf B$
satisfy ${\bf Q}^t\bf GQ=\bf G$ and ${\bf Q}^t\bf BQ=\bf B,$
then ${\bf p}_{L}$ and ${\bf p}_{R}$ are rotated by ${\bf Q}^*,$  with the
appropriate transformation of the momenta $\bf m,$ provided that  $\bf V$ does
not have entries in the rotated directions of the $E_8\times E_8$ lattice.
Moreover if one decomposes the Wilson line as $\bf W={\bf A}+\bf a$ where ${\bf
a}$ is the unrotated piece under the action of $\bf M,$
and ${\bf A} $ is perpendicular to ${\bf a} $ in the sense that ${\bf A}^t
{\bf C} {\bf a} \, = \, 0 $,
then it can be shown [\five] that ${\bf A}$ takes continuous values while $\bf
a$ is discrete. In this case the condition \back\ gives
$$\eqalign{{\bf M}{\bf A}-{\bf A}{\bf Q}=&{\bf 0}\cr
{\bf a}-{\bf aQ}=&{\bf V}\in Z,}\eqn\ten$$
and the action of the twist on the quantum numbers can be represented as
$${\bf u}\longrightarrow {\cal R}{\bf u};\quad {\cal R}=\pmatrix{
{\bf Q} & {\bf 0}& {\bf 0} \cr
{\bf \alpha} &{\bf Q}^* & \gamma \cr
{\bf a }({\bf 1}_d - {\bf Q}) & {\bf 0} &{\bf  M} \cr
},\eqn\mo$$
where $\alpha, \gamma$ are the matrices
$$\alpha = {1\over2} {\bf a}^t {\bf C} {\bf a} ({\bf 1}_d-{\bf Q}) + {1\over2}
({\bf 1}_d-{\bf Q}^*) {\bf a}^t{\bf C a}\in Z ,$$
$$\gamma =({\bf 1}_d-{\bf Q}^*){\bf a}^t{\bf C}\in Z. \eqn\twelve$$
Note that only those components of $\bf l$ that are not rotated by $\bf M$ get
shifted by ${\bf a} ({\bf 1}_d - {\bf Q}) {\bf n}$.

In order to study the duality symmetries in the presence of both discrete and
continuous Wilson lines, we employ the method used in [\eleven] and define the
new basis ${\bf u'}$
$${\bf u'}={\bf T}{\bf u}=\pmatrix{{\bf 1}_d&{\bf 0}&\bf 0\cr -{1\over2}{\bf
a}^t{\bf C}{\bf a}&{\bf 1}_d&-{\bf a}^t\bf C\cr
\bf a&\bf 0&{\bf 1}_{16}}{\bf u}\eqn\newbasis$$
In this basis the twist matrix is diagonal
$${\cal R}'=\pmatrix{
{\bf Q} & {\bf 0}& {\bf 0} \cr
{\bf 0}&{\bf Q}^* &{\bf 0}\cr
{\bf 0}&{\bf  0} &{\bf M }\cr
},\eqn\di$$
Due to the fact that ${\bf A}^t{\bf Ca}={\bf 0},$ the scaling dimension and
spin $H$ and $S$ take the form
$$H={1\over2}  {\bf u'}^t {\bf \Xi'} {\bf u'}, \qquad S={1\over2} {\bf u'}^t
\eta {\bf u'}\eqn\scale$$
where
$$\bf\Xi'=\pmatrix{
{1\over2}{\bf D'}{\bf G}^{-1} {\bf D'}^t &{\bf 1}_d+{1\over2}{\bf D'}{\bf
G}^{-1}&-{1\over2}{\bf D'}{\bf G}^{-1} {\bf A}^t {\bf C } \cr
 {\bf 1}_d+{1\over2}{\bf G}^{-1}{\bf D'}^t &{1\over2}{\bf G}^{-1}
&-{1\over2}{\bf G}^{-1} {\bf A}^t{\bf C} \cr
-{1\over2}{\bf C}{\bf A}{\bf G}^{-1}{\bf D}^t&-{1\over2}{\bf C}{\bf A}{\bf
G}^{-1} & {\bf C}+{1\over2}{\bf C}{\bf A}{\bf G}^{-1}{\bf A}^t{\bf
C}\cr}\eqn\nic$$
where ${\bf D'}=2\Big({\bf B}-{\bf G}-{1\over4}{\bf A}^t{\bf C}{\bf A}\Big)$.
The duality transformations of the orbifold are then given by those elements of
$O(d+16,6;Z)$ commuting with ${\cal R}'$ and satisfying
$${\bf T}^{-1}{\bf \Omega}^{-1}{\bf T}\in Z.\eqn\ol$$
The condition  \ol\  arising from the fact that the quantum numbers should
transform by an integer-valued transformation, constrains the parameters of
${\bf \Omega}$
and thus breaks the duality group to a subgroup whose form depends on the
choice of the discrete Wilson lines.

Having determined the duality group for the untwisted sector of orbifold
compactification in the presence of Wilson lines, we now turn to discuss the
symmetries of the twisted sectors. The twisted sectors are not sensitive to the
background fields unless the associated twist leaves invariant a particular
plane of the orbifold three complex planes. These twisted sectors are of
particular importance as they contribute to the moduli-dependent threshold
corrections to the gauge coupling constants [\twentyone-\twentythree].

Consider a $k$-twisted sector in a six-dimensional orbifold compactified
heterotic string theory, in which the associated twist $\theta^k$ represented
by ${\cal R}^k$ leaves a complex plane of the orbifold invariant.
This sector will have quantum numbers satisfying
$${\bf Q}^k{\bf n}={\bf n};\qquad {{\bf Q}^*}^k{\bf m}={\bf m};\qquad {\bf
M}^k{\bf l}={\bf l}.\eqn\hell$$
 Let $E_a, \, a = 1,2 $  be a set of  basis vectors for
  the  directions of the orbifold invariant under the action of of the twist
$\theta^k$ and ${\cal E}_\mu$, $\mu=1,\cdots ,d'$ be  a basis for the invariant
$E_8\times E_8$ directions.  Clearly these invariant directions are expressed
as some integral linear combinations of the orbifold lattice vectors $e_i$ and
the $E_8\times E_8$ lattice vectors $e_I.$
Then the twisted states will have a winding and momentum  vectors
$ L $ and $P$  given by
$$
L={\hat n}^{1}E_{1}+{\hat n}^{2}E_{2}, \, \, \, \,
P={\hat m}_1\tilde{E}_1+{\hat m}_2\tilde{E}_2
\eqn\as $$
where $\tilde{E}_1 , \tilde{E}_2 $ are  certain linear combinations
of the dual basis vectors $e^*_i $ with $ e^*_i\cdot e_j = \delta_{ij} $,
and ${\hat n}^{1}, {\hat n}^{2}, {\hat m}_1 , {\hat m}_2 $ are integers.
Note that  $\tilde{E}_a  $ are not necessarily orthogonal to
${E}_b $ so we define $\tilde{E}_{a} \cdot {E}_{b}
\, = \, {\alpha}_{ab} $.
Represent the quantum numbers  corresponding to the invariant
directions
by the matrices
$$\hat{\bf n}=\pmatrix {{\hat n}^{1}\cr {\hat n}^{2}};\quad \hat{\bf
m}=\pmatrix {{\hat m}_1\cr {\hat m}_2}; \quad
\hat {\bf l}=\pmatrix {{\hat l}_1\cr \vdots\cr {\hat l}_{d'}}\eqn\gif$$
and the background fields by
the $2\times 2$ matrices ${\bf G}_\perp$, ${\bf B}_\perp$ and the $d'\times 2$
matrix ${\bf A}_\perp$ where  ${\bf G}_\perp$ and $\bf B_\perp$ are the metric
and antisymmetric tensor of the invariant plane, ${\bf A}_\perp$ is the matrix
representing the continuous Wilson lines $A_{a }^{\mu}$. Clearly  ${\bf
G}_\perp,$ $\bf B_\perp$ and ${\bf A}_\perp$ are constructed from the original
$6\times 6$  matrices  $\bf G$ and $\bf B$ and from the $16\times 6$  matrix
$\bf  A$ respectively and satisfy
$${\bf Q}_\perp^t{\bf G}_\perp{\bf Q}_\perp={\bf G}_\perp,\quad {\bf
Q}_\perp^t{\bf B}_\perp{\bf Q}_\perp={\bf B}_\perp,\quad {\bf M}_\perp {\bf
A}_\perp={\bf A}_\perp {\bf Q}_\perp,\eqn\susy$$
where ${\bf Q}_\perp$ and ${\bf M}_\perp$ defines the action of the twist on
the directions $E_{a}$ and ${\cal E}_\mu$ respectively. Note also that ${\bf
A}_\perp$ will have non-vanishing components only in the rotated directions of
${\cal E}_\mu.$
Then  in the lattice basis the twisted sector will have, in matrix notation,
the following left and right moving momenta
$$\eqalign{{\bf P}_L=&\Big({{\bf \alpha} \hat {\bf m}\over2} + ({\bf G}_\perp
-{ \bf B}_\perp-{1\over4} {\bf A}_\perp^t {\bf C}_\perp{\bf A}_\perp)\hat {\bf
n} - {1\over2}{\bf A}_\perp^t{\bf  C}_\perp \hat{\bf l},\  \hat {\bf l}+{\bf
A}_\perp \hat {\bf n}\Big)=\Big({\bf p}_L,\ \tilde{{\bf p}}_L\Big),\cr
{\bf P}_R=&
\Big({{\bf \alpha} \hat {\bf m}\over2} - ({\bf G}_\perp +{ \bf
B}_\perp+{1\over4} {\bf A}_\perp^t {\bf C}_\perp{\bf A}_\perp)\hat {\bf n} -
{1\over2}{\bf A}_\perp^t{\bf  C}_\perp \hat{\bf l},\  {\bf 0}\Big)=\Big({\bf
p}_R,\ \bf 0\Big),}\eqn\per$$
 where $\alpha$ is the matrix $\alpha_{ab}$ and ${\bf C}_\perp$ is a $d'\times
d'$ representing the Cartan matrix of the directions  ${\cal E}_\mu$.
The world sheet energy and momentum of these twisted states now takes the form
$$H={1\over2} {\hat{\bf u}}^t {\bf \Xi}_\perp\hat{\bf u}_\perp,\qquad P
={1\over2} {\hat{\bf u}}^t \eta \hat {\bf u},\quad \hat{\bf
u}=\pmatrix{\hat{\bf n}\cr \bf\alpha\hat{\bf m}\cr \hat{\bf l} } \eqn\lon$$
where
$\eta_\perp$ and $\bf\Xi_\perp$ are  defined by  \nicosia, with a two
dimensional identity and the replacement of $\bf G$, $\bf B,$ $\bf A$ and $\bf
C$ by $\bf G_\perp$, $\bf B_\perp,$ $\bf A_\perp$ and ${\bf C}_\perp$
respectively. As an illustrative example, consider the orbifold
${\bf Z}_6-II$, with the twist defined by $\theta=(2,1,-3)/6$ and an
 $SU(6)\times SU(2)$ lattice.\footnote*{ the notation $(\zeta_1 , \zeta_2,
\zeta_3)$ is such
that the action of $\theta $ in the complex basis is \break
$({e}^{2 \pi {\rm i} \zeta_1},
{e}^{2 \pi {\rm i} \zeta_2}, {e}^{2 \pi {\rm i} \zeta_3}) $.}
The matrix $Q$
defining the twist action on the quantum numbers is given by
$$Q=\pmatrix{0&0&0&0&-1&0\cr
1&0&0&0&-1&0\cr 0&1&0&0&-1&0\cr 0&0&1&0&-1&0\cr 0&0&0&1&-1&0\cr 0&0&0&0&0&-1}
 \eqn\turkey $$
Consider the  $\theta^2$-twisted sector which leave the
the first ${\bf Z}_2$ complex plane invariant.
The invariant direction of the lattice and its dual under the $\theta^2$ action
are given by
$$\eqalign { E_{1}=&e_1+e_3+e_5,\quad E_{2}=e_6,\cr
\tilde{E}_1=&e^1-e^2+e^3-e^4+e^5,\quad \tilde{E}_2=e^6,\cr
\hbox{with}\  {\alpha}_{ab}  =&\pmatrix{3&0\cr 0&1}.} \eqn\heyou $$
The matrices ${\bf G}_{\perp}$ and ${\bf B}_{\perp}$ which are defined in terms
of the $E_{a}$ can be easily extracted from $\bf G$ and $\bf B,$ defining the
background of the six-dimensional orbifold.
Clearly the form of the Wilson lines ${\bf A}_\perp$ and the Cartan metric
${\bf C}_{\perp}$ will depend on the matrix ${\bf M},$
and its invariant directions.

Next we study the duality symmetry of these twisted sectors. First let us
consider the case where the invariant plane is a ${\bf Z}_2$-plane with
continuous Wilson lines only. Moreover let us consider the situation where
$\bf\alpha={\bf 1}_2,$ $i.e.,$ the invariant directions lie entirely in a
two dimensional sublattice of the orbifold six-dimensional lattice.  We look
for symmetries which leave both the spectrum of the twisted sector  and
$\hat{\bf l}+{{\bf A}_\perp}\hat{\bf n}$ invariant. This is sufficient to
ensure the invariance of the threshold corrections.
The most general form of the duality transformations with these requirements is
given by
$${\bf\Omega}_{{\bf Z}_2}=\pmatrix{{\bf F}&{\bf 0}&{\bf 0}\cr {\bf
F}^*\Big({\bf J}-{1\over2}{\bf V}_1^t{\bf C}_\perp{\bf V}_1\Big)&{\bf
F}^*&-{\bf F}^*{\bf V}^t_1{\bf C}_\perp&\cr {\bf V}_1&{\bf 0}&{\bf 1}_{d'}
}\eqn\general$$
where $\bf F$ is an $SL(2, Z)$ matrix satisfying $\bf F\bf Q_{\perp}=\bf
Q_{\perp}\bf F$, $\bf J$ is any antisymmetric integer matrix and ${\bf V}_1$ is
an integer matrix satisfying
${\bf V}_1{\bf Q}_{\perp}={\bf M}_{\perp}{\bf V}_1.$
Under the action of ${\bf\Omega}_{{\bf Z}_2}$ the transformations of the
background field can be obtained from  \sand. However, a simpler method of
obtaining the transformation law on the moduli was given in [\nine]. There one
defines the projective coordinate ${\bf P}=\pmatrix {-{\bf C}_\perp{\bf
A}_\perp\cr {\bf D}_\perp}$, where ${\bf D}_\perp=2\Big({\bf B}_\perp- {\bf
G}_\perp -{1\over4} {\bf A}_\perp^t{\bf C}_\perp{\bf A}_\perp\Big).$ This
coordinate
transforms under the action of ${\bf\Omega}_{{\bf Z}_N}$ as
$$\eqalign{{\bf P}&\longrightarrow({\bf X}_1{\bf P}+{\bf X}_2)({\bf X}_3{\bf
P}+{\bf X}_4)^{-1},\cr
{\bf X}_1=&\pmatrix{{\bf 1}_{d'}&{\bf 0}\cr {\bf V}^t_1&{\bf F}^t},\quad
{\bf X}_2=
\pmatrix{-{\bf C}_\perp{\bf V}_1{\bf F}^{-1}\cr -({\bf J}+
{1\over2}{\bf V}_1^t{\bf C}_\perp{\bf V}_1){\bf F}^{-1}},\cr
{\bf X}_3=&
\pmatrix{{\bf 0}& {\bf 0}},\quad
{\bf X}_4={\bf F}^{-1}.} \eqn\simpler$$
In particular, \simpler\ gives for the Wilson lines,
$${\bf A}_{\perp}\longrightarrow {\bf A}_{\perp}{\bf F}+{\bf V}_1.\eqn\wil$$
Clearly, the duality group defined by \general\ contains the $U$-duality as a
subgroup. This is the subgroup defined by
$${\bf\Omega}_{U}=\pmatrix{{\bf F}&{\bf 0}&{\bf 0}\cr {\bf 0}&{\bf F}^*&{\bf
0}&\cr {\bf 0}&{\bf 0}&{\bf 1}_{d'}  }\eqn\u$$
with $det {\bf F}=1$. Its action on the quantum numbers and the background
fields is given by
$$\eqalign{\hat{\bf n}&\longrightarrow {\bf F}^{-1}\hat{\bf n},\cr
\hat{\bf m}&\longrightarrow {\bf F}^t\hat{\bf m},\cr
\hat{\bf l}&\longrightarrow\hat{\bf l},\cr
({\bf G}_\perp\pm{\bf  B}_\perp)&\longrightarrow {\bf F}^t({\bf G}_\perp\pm{\bf
B}_\perp){\bf F},\cr
 {\bf A}_\perp&\longrightarrow {\bf A}_\perp{\bf F}.}\eqn\udual$$
Another symmetry  is obtained by setting $\bf F=\pmatrix{1&0\cr 0&-1}.$  This
symmetry, for the case where the Wilson line has two gauge indices and the rest
vanishing, acts on the complex moduli defined in [\nine] by
$$T\longrightarrow {\bar T}, \quad U\longrightarrow {\bar U}, \quad
B\longrightarrow  {\bar C}, \quad C\longrightarrow {\bar B}\eqn\must$$

Another subgroup of \general\ is   the axionic shift symmetry which is given by
$${\bf\Omega}=\pmatrix{{\bf 1}_2&{\bf 0}&{\bf 0}\cr {\bf J}&{\bf 1}_2&{\bf
0}&\cr {\bf 0}&{\bf 0}&{\bf 1}_{d'} }\eqn\ax$$
This subgroup acts on the quantum numbers and moduli as
$$\eqalign{\hat{\bf n}&\longrightarrow \hat{\bf n}\cr
\hat{\bf m}&\longrightarrow \hat{\bf m}-{\bf J}\hat{\bf n},\cr
\hat{\bf l}&\longrightarrow\hat{\bf l},\cr
{\bf G}_\perp&\longrightarrow {\bf G}_\perp,\cr
{\bf B}_\perp&\longrightarrow{\bf B}_\perp- {{\bf J}\over2},\cr
{\bf A}_\perp&\longrightarrow {\bf A}_\perp.}\eqn\ax$$
This symmetry, is a subgroup of the $T$-duality symmetry. The other elements of
the $T$ duality do not leave $\hat{\bf l}+{\bf A}_\perp\hat{\bf n}$ invariant
as it involves the interchange of windings and momenta as well as the mixing of
${\bf A}_\perp$ with ${\bf G}_\perp$ and ${\bf B}_\perp$.
Therefore we conclude that the symmetries of the threshold corrections to the
gauge coupling constants coming from a ${\bf Z}_2$ invariant plane are those
which transform the continuous Wilson lines by an $SL(2, Z)$ rotation plus a
shift.
If we consider a ${\bf Z}_2$-plane with $ \alpha\not= {\bf 1}_2$, then the
duality symmetry is given by \general\ but with the additional condition
$$\eqalign{{\alpha}^{-1}{\bf F}^t\alpha& \in Z,\cr
\alpha^{-1}{\bf V}_1^t{\bf C}_\perp&\in Z\cr
{\alpha}^{-1}\Big({\bf J}+{1\over2}{\bf V}_1^t{\bf C}_\perp{\bf V}_1\Big)&\in
Z.}\eqn\moh$$
Such conditions arise from the fact the quantum numbers should transform as
integers.

The above considerations can be extended to a ${\bf Z}_N $ plane with
$N \neq 2 $. Here the duality symmetry is given by
$${\bf\Omega}_{{\bf Z}_N}=\pmatrix{{\bf 1}_2 &{\bf 0}&{\bf 0}\cr \Big({\bf
J}-{1\over2}{\bf V}^{'t}_{1}{\bf C}_\perp{\bf V}^{'}_{1}\Big)&{\bf 1}_{2}&-{\bf
V}^{'t}_{1}{\bf C}_\perp&\cr {\bf V}^{'}_{1}&{\bf 0}&{\bf 1}_{d'} }\eqn\gen$$
with the condition
${\bf V}^{'}_1{\bf Q}_{\perp}={\bf M}_{\perp}{\bf V}^{'}_1$ coming from the
fact that  ${\bf\Omega}_{{\bf Z}_N} $ has to commute with the ${\bf Z}_{N} $
twist. Also in the case when ${\alpha }\neq  {\bf 1}_2 $, a further breaking of
the symmetry occurs due to the conditions
$$\eqalign{
\alpha^{-1}{\bf V}^{'t}_1{\bf C}_\perp&\in Z\cr
{\alpha}^{-1}\Big({\bf J}+{1\over2}{\bf V}^{'t}_1 {\bf C}_\perp{\bf
V}^{'}_1\Big)&\in Z.}\eqn\mugs$$

If  we now allow the presence of discrete Wilson lines $\bf a$,
the twisted sectors with invariant planes have left and right moving momenta of
the form
\per\ but with $\bf\hat n,$ $\bf\hat m$ and $\bf\hat l$ replaced by the
$\bf\hat n',$ $\bf\hat m'$ and $\bf\hat l'$ where the latter quantum numbers
are the independent entries of the solutions of
$${\bf Q}^k{\bf n'}={\bf n'};\qquad {{\bf Q}^*}^k{\bf m'}={\bf m'};\qquad {\bf
M}^k{\bf l'}={\bf l'}.\eqn\he$$
where $({\bf n'}, {\bf m'}, {\bf l'})$ are the basis defined in \newbasis.
If we write
$$\hat {\bf u'}=\pmatrix{\hat {\bf n'}\cr \hat{\bf m'}\cr \hat{\bf  l'}}={\cal
T}\hat {\bf u}
\eqn\hungry$$
where ${\cal T}$ depends on the choice of discrete Wilson lines of the orbifold
model considered, then the duality symmetry of the threshold corrections in the
presence of discrete Wilson lines are those symmetries obtained in the absence
of discrete Wilson lines with the additional constraint
$${\cal T}^{-1}{\bf\Omega}_{{\bf Z}_N}{\cal T}\in Z.\eqn\fin$$
The condition \fin\ is the statement that the duality transformation should act
on the quantum numbers $\hat {\bf u}$ by an integer-valued matrix.
When ${\bf M}  = {\bf I } $ (i.e. in the absence of  continuous Wilson lines ),
the form of ${\cal T} $ for  all  $ {\bf Z}_N $
            orbifolds was derived in [\eleven  ] .

In summary, we have studied the duality symmetry of orbifold compactification
in the presence of a general background. We have
derived the duality
symmetries of the threshold corrections to the gauge coupling constants in the
presence of continuous Wilson lines (Wilson line moduli), and demonstrated
that these symmetries
are  broken  to a subgroup if one includes discrete Wilson lines.
The threshold corrections in the presence of Wilson lines could be of
fundamental phenomenological importance in the study of the gauge couplings
unification. In the absence of Wilson lines, the expectation values of the
moduli which gives rise to the  unification scale of the gauge coupling
constant are not in agreement with those obtained through the minimization
of a possible non-perturbative superpotential. The dependence of the
threshold corrections on the Wilson lines moduli gives more degrees of
freedom which might resolve this discrepancy.
\vskip0.5in
\centerline{\bf{ACKNOWLEDGEMENT}}
We would like to thank S. Stieberger for useful conversations.
This work is supported in part by S.E.R.C. and the work of S.T  is
supported by the Royal Society.
\refout
\end